\documentclass[11pt]{article}

\usepackage[letterpaper,margin=1in]{geometry}
\usepackage[T1]{fontenc}
\usepackage[utf8]{inputenc}
\usepackage{lmodern}
\usepackage{microtype}
\usepackage{amsmath,amssymb,amsthm}
\usepackage{enumitem}
\usepackage[hidelinks]{hyperref}

\newtheorem{definition}{Definition}
\newtheorem{question}{Question}
\newtheorem{conjecture}{Conjecture}

\newcommand{\SigmaPlus}{\Sigma^{+}}
\newcommand{\Lang}{L}
\newcommand{\ID}{\operatorname{ID}}
\newcommand{\varnum}{\#\operatorname{var}}

\title{Compute Inclusion Depth of a Pattern}
\author{Wei Luo\\
School of Information Technology, Deakin University\\
\texttt{wei.luo@deakin.edu.au}}
\date{Author-prepared version, 2026}

\begin{document}
\maketitle

\begin{abstract}
This note formulates the problem of computing the inclusion depth of a pattern language. Inclusion depth is intended to capture the mind-change complexity of pattern identification problems. The central question is whether inclusion depth is computable, and in particular whether it admits a simple combinatorial characterisation in terms of the length of a pattern and the number of distinct variables occurring in it.
\end{abstract}

\paragraph{Publication note.}
This is an author version of an open problem note that appeared in \emph{Learning Theory: 18th Annual Conference on Learning Theory, COLT 2005}, Lecture Notes in Computer Science 3559, pp.~689--690. The formatting here is intentionally generic and does not use the original publisher style.  The original version is available at \url{https://link.springer.com/chapter/10.1007/11503415_48}.

\section{Problem Description}

We define a concept of \emph{inclusion depth} (Definition~\ref{def:inclusion-depth}) to capture mind-change complexity~\cite{freivalds1993role,ambainis1999ordinal} of pattern identification problems~\cite{angluin1980finding}. Our basic question is whether the inclusion depth for any pattern is computable. We conjecture a combinatorial characterisation that, if true, leads to a linear-time algorithm to compute inclusion depth.

Let $X$ be a set of variables, for example $x_1,x_2,\ldots$, and let $\Sigma$ be a finite alphabet containing at least two symbols, for example $\{0,1\}$. A \emph{pattern}, denoted by $p,q$, and so on, is a finite non-empty sequence over $X \cup \Sigma$. The language of a pattern $p$ with alphabet $\Sigma$, denoted by $\Lang_\Sigma(p)$, is the set of ground strings that are consequences of $p$ by substituting each variable in $p$ with a string in $\SigmaPlus$. For example, if $\Sigma=\{0,1\}$, then the strings $010$ and $10110$ are in $\Lang_\Sigma(x_1x_2x_1)$, but $1010 \notin \Lang_\Sigma(x_1x_2x_1)$.

\begin{definition}[Inclusion depth]\label{def:inclusion-depth}
The \emph{inclusion depth} of a pattern $p$ with alphabet $\Sigma$, denoted by $\ID_\Sigma(p)$, is the length of the longest strict inclusion chain connecting the language of the universal pattern $\Lang_\Sigma(x_1)$ and the language $\Lang_\Sigma(p)$.
\end{definition}

For example, if $p=0x_1 1$ and $\Sigma=\{0,1\}$, then
\begin{equation*}
\Lang_\Sigma(x_1) \supset \Lang_\Sigma(x_1x_2) \supset \Lang_\Sigma(x_1 1) \supset \Lang_\Sigma(x_1x_2 1) \supset \Lang_\Sigma(0x_1 1)
\end{equation*}
is an inclusion chain connecting $\Lang_\Sigma(x_1)$ and $\Lang_\Sigma(0x_1 1)$. It is routine, though tedious, to verify that there exists no longer inclusion chain connecting the two. Thus the inclusion depth $\ID_\Sigma(0x_1 1)$ is equal to $4$.

\begin{question}\label{q:computable}
Is there an algorithm to compute $\ID_\Sigma(p)$ for any pattern $p$ and any alphabet $\Sigma$? If yes, is there a polynomial-time algorithm?
\end{question}

\begin{question}\label{q:formula}
Is it true that, for every alphabet $\Sigma$ with at least two symbols,
\begin{equation}\label{eq:main-formula}
\ID_\Sigma(p) = 2|p| - \varnum(p) - 1,
\end{equation}
where $|p|$ is the length, namely the number of variables and constants in $p$, and $\varnum(p)$ is the number of distinct variables in $p$?
\end{question}

If the answer to Question~\ref{q:formula} is yes, then we have a linear-time algorithm that computes inclusion depth using Equation~\eqref{eq:main-formula}, and hence a positive answer to Question~\ref{q:computable}.

\section{Motivation}

Inclusion depth captures the mind-change complexity of the pattern identification problem given some initial evidence. If there exists an algorithm to compute the inclusion depth of a pattern, then we can use it to compute the mind-change bound\footnote{In problems of learning languages with positive data, the mind-change bound~\cite{ambainis1999ordinal} measures the worst-case number of mind changes a learner has to make before it converges to the correct answer in the sense defined by Gold~\cite{gold1967language}.} of a pattern identification problem given some initial evidence. Moreover, we can construct a uniformly mind-change optimal learner~\cite{luo2005mind}.

\section{Partial Solution and Difficulties}

It is known that pattern inclusion is undecidable~\cite{jiang1993inclusion}. However, to compute the inclusion depth, we may not need to decide inclusion for arbitrary pairs of patterns.

Intuitively, patterns with longer length are more constrained, and so are patterns with fewer distinct variables. From observing some shorter examples, this suggests a conjecture that for a pattern $p$, the inclusion depth $\ID(p)$ is related to the length $|p|$ and the number of distinct variables $\varnum(p)$ by Equation~\eqref{eq:main-formula}. To prove this equation, it suffices to establish the following conjecture.

\begin{conjecture}\label{conj:monotone-measure}
Let $p$ and $q$ be two patterns. If $\Lang(p) \subset \Lang(q)$, then
\begin{equation}\label{eq:monotone-measure}
2|p| - \varnum(p) > 2|q| - \varnum(q).
\end{equation}
\end{conjecture}

To see why the reference to the alphabet $\Sigma$ is dropped, note that if $\Lang_\Sigma(p) \subseteq \Lang_\Sigma(q)$, then for every $\Sigma' \subseteq \Sigma$, we have $\Lang_{\Sigma'}(p) \subseteq \Lang_{\Sigma'}(q)$. Therefore, we need to consider only the case in which $\Sigma$ contains exactly two symbols.

Exhaustive computation by the author's program showed that Equation~\eqref{eq:monotone-measure} holds for patterns of length less than or equal to $7$; computation for longer patterns was computationally impractical. Therefore, if two patterns $p$ and $q$ form a counterexample to Equation~\eqref{eq:monotone-measure}, one of them must be longer than $7$.

\bibliographystyle{plain}
\bibliography{references}

@article{ambainis1999ordinal,
  author  = {Ambainis, Andris and Jain, Sanjay and Sharma, Arun},
  title   = {Ordinal Mind Change Complexity of Language Identification},
  journal = {Theoretical Computer Science},
  volume  = {220},
  number  = {2},
  pages   = {323--343},
  year    = {1999}
}

@article{angluin1980finding,
  author  = {Angluin, Dana},
  title   = {Finding Patterns Common to a Set of Strings},
  journal = {Journal of Computer and System Sciences},
  volume  = {21},
  number  = {1},
  pages   = {46--62},
  year    = {1980}
}

@article{freivalds1993role,
  author  = {Freivalds, Rusins and Smith, Carl H.},
  title   = {On the Role of Procrastination in Machine Learning},
  journal = {Information and Computation},
  volume  = {107},
  number  = {2},
  pages   = {237--271},
  year    = {1993}
}

@article{gold1967language,
  author  = {Gold, E. Mark},
  title   = {Language Identification in the Limit},
  journal = {Information and Control},
  volume  = {10},
  number  = {5},
  pages   = {447--474},
  year    = {1967}
}

@inproceedings{jiang1993inclusion,
  author    = {Jiang, Tao and Salomaa, Arto and Salomaa, Kai and Yu, Sheng},
  title     = {Inclusion Is Undecidable for Pattern Languages},
  booktitle = {Automata, Languages and Programming: 20th International Colloquium, ICALP 1993},
  editor    = {Lingas, Andrzej and Karlsson, Rolf G. and Carlsson, Svante},
  series    = {Lecture Notes in Computer Science},
  volume    = {700},
  pages     = {301--312},
  publisher = {Springer},
  year      = {1993}
}

@inproceedings{luo2005mind,
  author    = {Luo, Wei and Schulte, Oliver},
  title     = {Mind Change Efficient Learning},
  booktitle = {Learning Theory: 18th Annual Conference on Learning Theory, COLT 2005, Bertinoro, Italy, June 27--30, 2005, Proceedings},
  editor    = {Auer, Peter and Meir, Ron},
  series    = {Lecture Notes in Computer Science},
  volume    = {3559},
  pages     = {398--412},
  publisher = {Springer},
  year      = {2005}
}

\end{document}